\documentclass[pra,aps,twocolumn,showpacs,groupedaddress,superscriptaddress]{revtex4}

\usepackage[latin1]{inputenc}
\usepackage[english]{babel}
\usepackage{graphicx}
\begin{document}

\title{Zeno dynamics in wave-packet diffraction spreading}

\author{Miguel A. Porras}
\affiliation{Departamento de F\'{\i}sica Aplicada, Universidad
Polit\'{e}cnica de Madrid, Rios Rosas 21, 28003 Madrid, Spain}

\author{Alfredo Luis}
\affiliation{Departamento de \'{O}ptica, Facultad de Ciencias
F\'{\i}sicas, Universidad Complutense, 28040 Madrid, Spain}

\author{Isabel Gonzalo}\email{igonzalo@fis.ucm.es}
\affiliation{Departamento de \'{O}ptica, Facultad de Ciencias
F\'{\i}sicas, Universidad Complutense, 28040 Madrid, Spain}

\author{\'{A}ngel S. Sanz}
\affiliation{Instituto de F\'{\i}sica Fundamental -- CSIC, Serrano 123,
28006 Madrid, Spain}

\begin{abstract}
We analyze a simple and feasible practical scheme displaying Zeno,
anti-Zeno, and inverse-Zeno effects in the observation of wave-packet
spreading caused by free evolution. The scheme is valid
both in spatial diffraction of classical optical waves and in time
diffraction of a quantum wave packet. In the optical realization,
diffraction spreading is observed by placing slits between a
light source and a light-power detector. We show that the
occurrence of Zeno or anti-Zeno effects depends just on the
frequency of observations between the source and detector. These
effects are seen to be related to the diffraction mode theory in
Fabry-Perot resonators.
\end{abstract}


\maketitle

\section{Introduction}

We owe to quantum mechanics the most subtle reflections on the
subject-object relation in observation processes. This includes
the unavoidable perturbation of the observed system, which
acquires the form of  Zeno-type effects when  its evolution is
repeatedly monitored along time \cite{Misra77,Khalfin57,Itano90}.
Zeno effects can be found at least in three forms: proper Zeno
effect (the evolution  decay is slowed down)
\cite{Misra77,Khalfin57,Itano90,Muga08}, anti-Zeno effect (the
evolution is speeded up)
\cite{Lane83,Schieve89,Kofman96,Alfredo98,Wilkinson97,Fischer01,Dreisow08,
Lizuain10}, and inverse-Zeno effect (the evolution is guided by
gradually changing measurements)
\cite{Kitano97,Yamane01,Aharonov80, Altenmuller93}. Remarkably,
the Zeno effects have crossed the quantum-classical border
\cite{Kitano97,Yamane01,cZ,LonghiPRL,LonghiOE}.

In this work we present an example of such Zeno effects by means
of an extremely simple and practical scheme. Any Zeno scheme
requires  a clear identification of two ingredients: the observed
dynamics and the measurement performed to observe  such a
dynamics.

Here the evolution to be observed  is the free diffraction
spreading of a wave packet in two different physical systems: (i)
a quantum free particle and (ii) classical light diffraction in
the Fresnel regime. The equivalence of these systems arises from
the fact that the dynamical evolution of both is ruled by the same
equation, namely the Schr\"odinger equation. In fact, free
particle evolution has been referred to as \textit{time
diffraction} in the literature (see, for example, \cite{dit}).
Time diffraction lowers the probability of finding the particle in
its initial region after any lapse of time. In classical wave
optics, diffraction lowers the light power reaching a distant,
finite detector from a finite source (we consider source and
detector of the same size).

We propose that the diffraction can be observed by inserting
intermediate slits between the source and the detector. These
slits play the role of measurements, since they make it possible to
detect the amount of power in the finite region they define. In other
words, these slits monitor whether spreading has already taken
place or not. This can be therefore considered a {\em bona fide}
Zeno scheme. The idea is to study how the light power reaching the
detector depends on the number of slits introduced and other
relevant parameters, such as source-detector distance. To be
convinced that this is worth investigating,
one can ask whether placing a slit midway between
two other slits will decrease the light power reaching the
detector (anti-Zeno effect) or otherwise will increase it (Zeno
effect). In time diffraction, the particle is prepared somewhere
within a finite spatial region, and the position is periodically
measured  to monitor whether the particle continues in the initial
region.

The merits of the  aforementioned scheme are, among others (for
definiteness we  mainly focus on the classical wave optics
realization), the following.
\begin{enumerate}
\item[(i)] There is a full equivalence between the quantum and classical-optics
versions. Classical versions of Zeno effects are welcome since they allow
us to better understand both classical wave optics and quantum mechanics.

\item[(ii)]
We find striking parallels with the diffraction modes of a Fabry-Perot resonator. They
were actually introduced as the waves shaped by repeated diffraction in the
finite-size resonator mirrors \cite{FL}. Moreover, this equivalence can be extended
to other physical processes, such as pulse compression by saturable absorbers,
for example.

\item[(iii)] Zeno, anti-Zeno, and inverse-Zeno effects can occur in the same
scheme by simply varying the relative positions of the source, intermediate slits, and
detector (or equivalently, by changing the slit width and light wavelength).

\item[(iv)] Zeno, anti-Zeno, and inverse-Zeno effects are clearly perceptible even
after a very small number of measurements.

\item[(v)] The classical-optics version has an extremely simple experimental
implementation, accessible even to undergraduate labs.

\item[(vi)] Contrary to the more standard Zeno effect, in our case the measurement does
not project the system into the initial state. The measurement
determines the total light power within the slit, or the
probability of finding the quantum particle in that region, but
otherwise the wave and particle are free to evolve respecting this
confinement. This is, in other words, an example of Zeno dynamics
\cite{Fakki,rev}.
\end{enumerate}

The organization of this work is as follows. In Sec.~\ref{NUM} we
present the results for the Zeno effects obtained from numerical
simulations. Their theoretical interpretation  is developed in
Sec.~\ref{THEORY}. The main conclusions are extracted in
Sec.~\ref{conc}.

\section{Zeno, anti-Zeno and inverse-Zeno effects in wave-packet spreading}\label{NUM}

\begin{figure}
\begin{center}
\includegraphics[width=6cm]{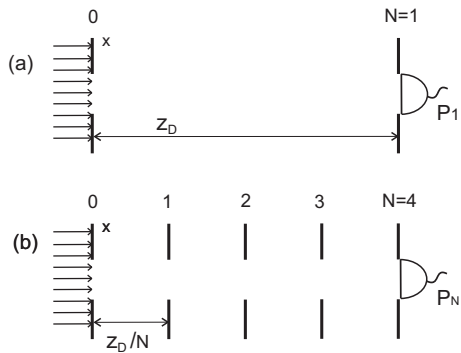}
\end{center}
\caption{\label{fig1} (a) Unobserved diffraction. (b) Observed diffraction.}
\end{figure}

Diffraction of the wave packet of envelope $\psi (x)$ by a slit of half-width $a$
can be described by the Fresnel diffraction integral,
which is conveniently written in the form
\begin{equation}\label{FRESNEL}
\psi_1(\xi)=\frac{1}{\sqrt{2\pi i\zeta}}\int_{-1}^{1} d\xi'\psi
(\xi')\exp\left[\frac{i}{2\zeta}(\xi'-\xi)^2\right]\, ,
\end{equation}
where $\xi= x/a$ is the spatial coordinate in units of $a$, and
$\zeta=z/(ka^2)$ is the propagation distance $z$ in units of the
diffraction distance $ka^2$, $k=2\pi/\lambda$ being the wave
number and $\lambda$ the  wavelength. Equation (\ref{FRESNEL})
describes also the spreading of the quantum wave packet of a
particle $\psi(x)$ with center of mass at rest confined in
$[-a,a]$ if $\zeta=\hbar t/(ma^2)$ is the time $t$ in units of the
diffraction time $ma^2/\hbar$.

\begin{figure}
\begin{center}
\includegraphics[width=4.1cm]{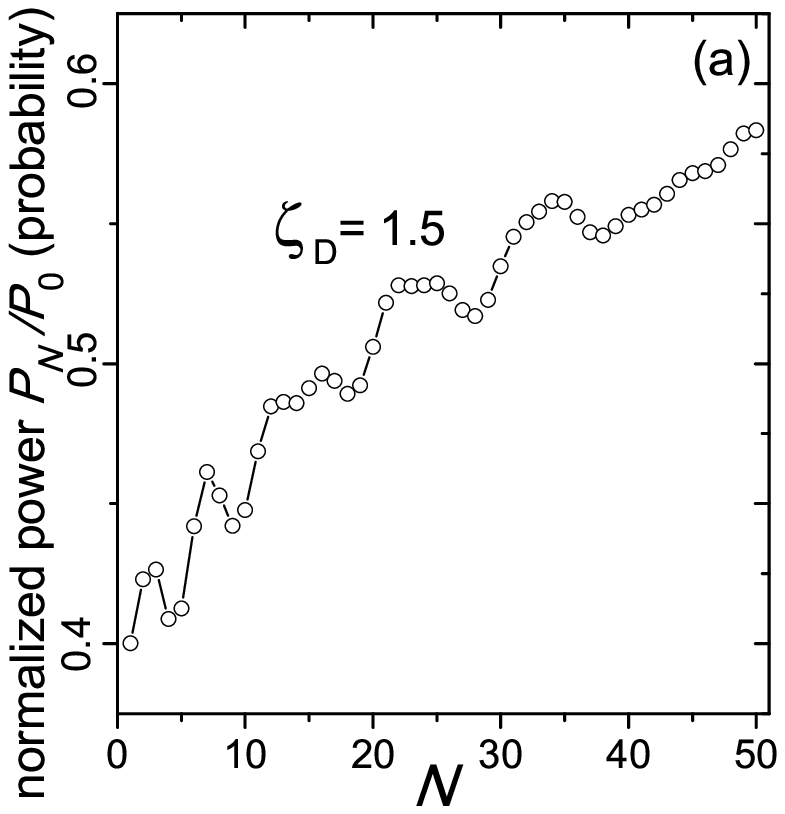}\hspace*{0.2cm}
\includegraphics[width=4cm]{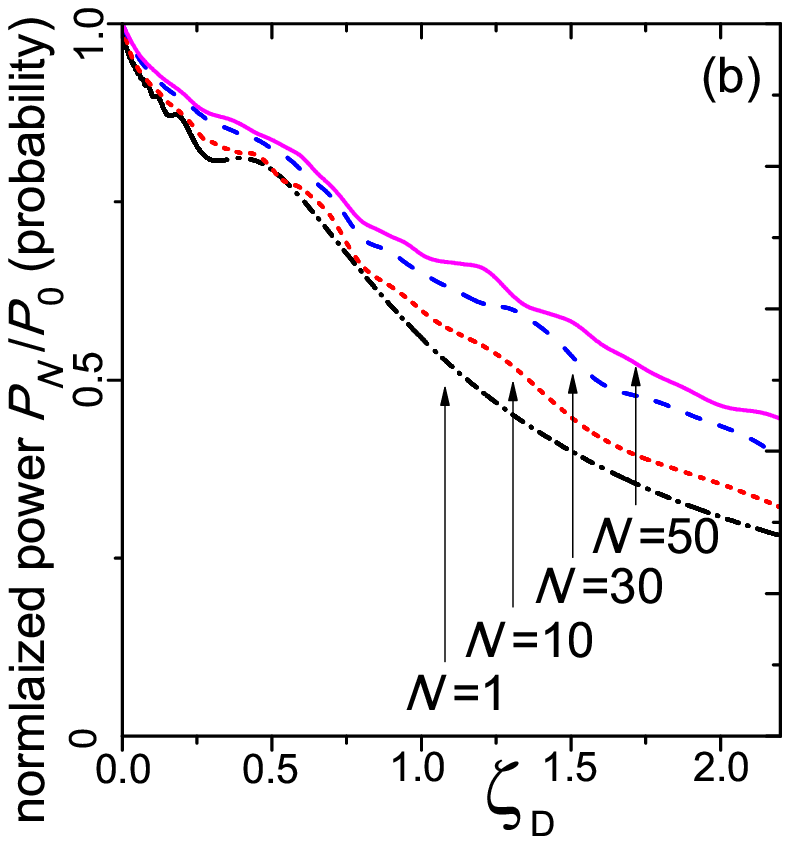}
\end{center}
\caption{\label{fig2}  (Color online) Normalized power $P_N/P_0$ in the detector
slit (a) as a function of the number of slits
 for a fixed source-detector distance $\zeta_{\rm D}$, and (b) as a
 function of the distance $\zeta_{\rm D}$ for a few,
 increasing number of slits within $\zeta_{\rm D}$.}
\end{figure}

In our numerical simulations, a uniform plane wave illuminates the
slit (Fig.~\ref{fig1}).  The detector then measures the light
power going through a slit of equal size placed at a fixed
distance $\zeta_{\rm D}$,  both when no slits are inserted
(unobserved diffraction  case) and when a number of intermediate
slits are evenly inserted (observed diffraction  case). Given $N$,
$n=0,1,\dots N$ denote the source, intermediate, and detector
slits at positions $\zeta_n=(\zeta_{\rm D}/N)n$. The number of
intermediate slits is then $N-1$ (in particular, $N=1$ means no
intermediate slits). In order to evaluate the field $\psi_N(\xi)$
at the detector slit and the captured power,
\begin{equation}
P_N=\int_{-1}^1 |\psi_N(\xi)|^2 d\xi\, ,
\end{equation}
we make use of Eq.~(\ref{FRESNEL}) $N$ times with
$\zeta=\zeta_{\rm D}/N$, starting with $\psi(\xi)={\rm const.}$ in
$[-1,+1]$, to find the field in each intermediate slit from the
preceding one until the detector slit. Figure \ref{fig2}(a)
represents the power $P_N$ as a function of $N$, and shows that
the detected power increases, though not monotonically, as more
and more intermediate slits are inserted, approaching the power on
the source slit, $P_0$, in the limit of large $N$. For $N=50$, for
example, we get about 50\% of power increase with respect to the
case of no intermediate slits. Figure \ref{fig2}(b) shows that the
eventual growth with large $N$ holds for all values of $\zeta_{\rm
D}$. The detected power without intermediate slits (lower curve)
is lower than the detected power with a large enough number of
inserted slits (upper curves). Translated  into a quantum
mechanics  language, we can affirm that a particle prepared in a
localized state is more likely to preserve its localization if
this property is checked a large enough number of times, since
$P_N$ represents the probability of finding the particle in the
localized region after $N$ measurements.

\begin{figure}[!]
\begin{center}
\includegraphics[width=5.5cm]{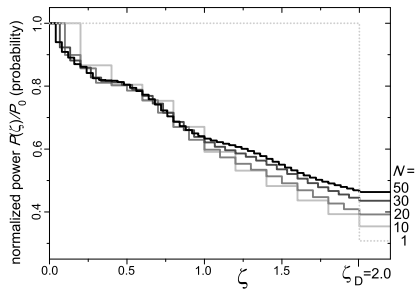}
\end{center}
\caption{\label{fig3} Normalized power $P(\zeta)/P_0$, where
$P(\zeta)$ is computed according to Eq. (\ref{power}), as a
function of the propagation distance $\zeta$ for different number
of intermediate slits within $\zeta_{\rm D}=2.0$. Steps are at
$\zeta_n=n(\zeta_{\rm D}/N)$, $n=0,1\dots N$ for each $N$. The
lowest step of each staircase is the power that arrives at the
detector in each case.}
\end{figure}

Figure \ref{fig3} illustrates how this Zeno effect is forged as
light is diffracted more and more times in the intermediate slits.
The power at a distance $\zeta$ is computed as
\begin{equation}
\label{power}
P (\zeta) = \int_{-\infty}^\infty |\psi_\zeta (\xi)|^2 d\xi \, ,
\end{equation}
where $\psi_\zeta (\xi)$ represents the wave at plane $\zeta$. In
absence of intermediate slits, the beam power is preserved
(horizontal lighter line) until the beam impinges the detector
slit, where it loses an important fraction of its power (vertical
lighter line). As intermediate slits are inserted and their number
increases, the ``staircases'' of power due to diffraction losses at
each slit (darker lines) have more but so less steep steps that the
power reaches higher and higher values at the detector. The same
description applies to the probability of finding the particle
after checking more and more frequently its presence in a finite
region of space.

\begin{figure}[b]
\begin{center}
\includegraphics[width=4cm]{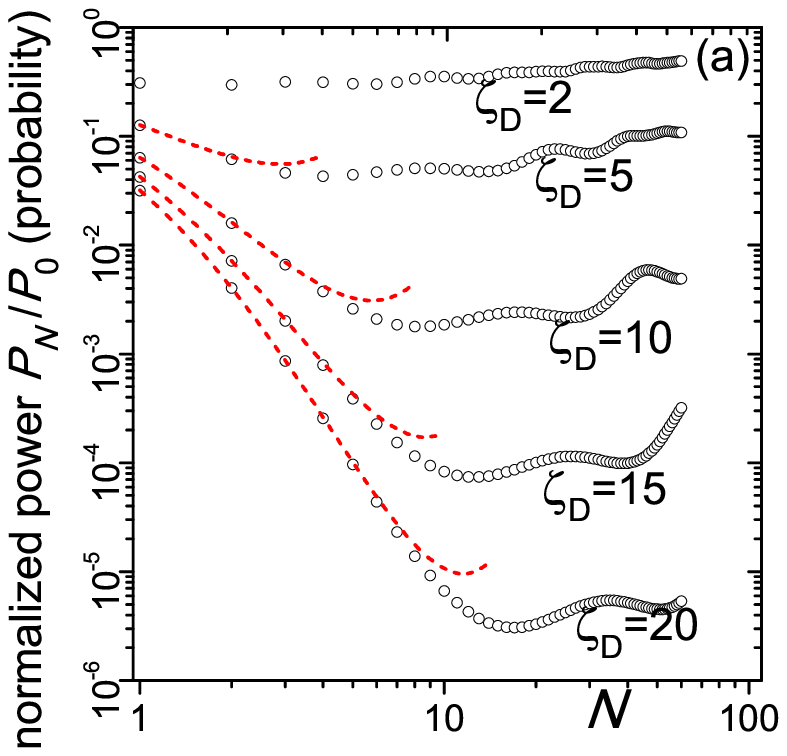}\includegraphics[width=3.8cm]{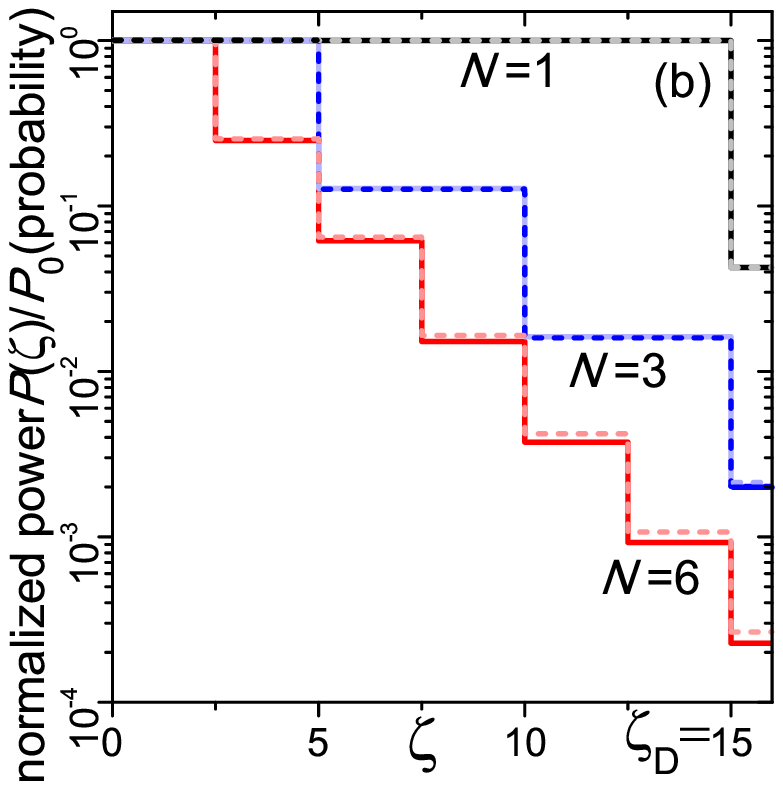}
\end{center}
\caption{\label{fig4} (Color online) (a) Open circles: Normalized power $P_N/P_0$ in the detector slit
as a function of the number of
intermediate slits for some given values of $\zeta_{\rm D}$. Dashed curves:
prediction of Eq.~(\ref{PROBANTI}), valid for $N\ll \zeta_{\rm
D}$. (b) Solid curves: Normalized power $P(\zeta)/P_0$ as a function of the
propagation distance $\zeta$ for different number of slits within
$\zeta_{\rm D}=15$. The lowest step of each staircase is the power
that arrives at the detector in each case. Dashed curves: the same
but using the approximate values $(2N/\pi \zeta_{\rm D})^n$,
$n=0,1\dots N$, at $\zeta_n=n(\zeta_{\rm D}/N)$.}
\end{figure}

As pointed out in the Introduction, a nonobvious fact is that the
power in the detector starts to increase from the very first
intermediate slit when the source-detector distance verifies
$\zeta_{\rm D}\lesssim 2$ [see,  for example, Fig.~\ref{fig2}(a)].
The opposite situation with $\zeta_{\rm D} \gtrsim 2$ is
illustrated in Fig.~\ref{fig4}. Initially, inserting  an
increasing number of slits results in a regular decrease of the
detected power [see Fig.~\ref{fig4}(a), open circles curves],
though the inclusion of further slits always reverses this trend.
This is the more intuitive situation in which the more the loss
events, the lower the power in the detector [Fig.~\ref{fig4}(b),
solid curves], and represents an anti-Zeno effect in wave-packet
spreading: Repeated observation of wave-packet spreading within a
certain distance or time $\zeta_{\rm D}$ enhances spreading
compared to the unobserved case. The anti-Zeno effect is
manifested when the observation  intervals, or slit-to-slit
distance,  are slightly larger than one diffraction length: as
seen in Fig.~\ref{fig4}(a), the detected power at $\zeta_{\rm
D}=5,10,15,20$ continuously decreases until
$N=4,8,12,16$ slits, respectively, are inserted [Fig.~\ref{fig4}(a)], which
yields a slit-to-slit distance $\zeta_{\rm D}/N\simeq1.25$ in all
four cases. A similar value is obtained in other cases. The value
$1.25$ times the diffraction length  can be then considered as a
Zeno distance (or time) in wave-packet spreading, which marks off
the transition from  a Zeno to  an anti-Zeno behavior.

\begin{figure}[t]
\begin{center}
\includegraphics[width=6cm]{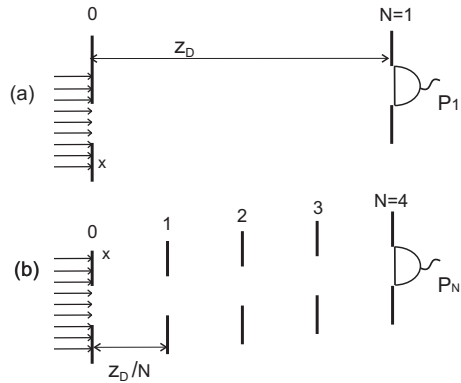}
\end{center}
\caption{\label{fig5} Scheme for the inverse-Zeno effect. (a)
Unobserved diffraction. (b) Observed diffraction.}
\end{figure}

As a variant of the Zeno effect, the inverse-Zeno effect can
also be observed in wave-packet spreading. The scheme of
Fig.~\ref{fig1} is now changed by the one displayed in
Fig.~\ref{fig5}, where the detector slit is laterally displaced so
that its bottom edge is at the same height as the top edge of the
source slit, and therefore they do not overlap. As in the
preceding cases, we insert equally spaced slits between the source
and the detector, but they are now gradually displaced in the
lateral direction by $\Delta \xi_n = n(2/N)$, $n=0,1 \dots N$, as
sketched in Fig.~\ref{fig5}. The inverse-Zeno effect occurs if the
light power reaching the detector in the observed diffraction case
is larger than in the unobserved diffraction case, meaning that
the wave has been guided by observation, at least partially, to
the non-overlapping detector.

\begin{figure}[h]
\begin{center}
\includegraphics[width=4.0cm]{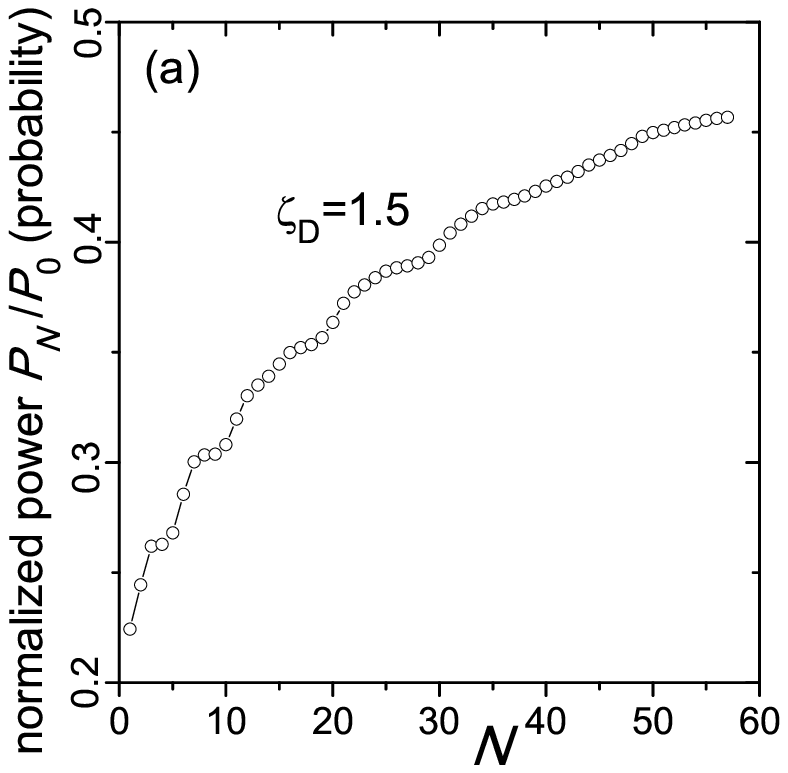}\includegraphics[width=3.9cm]{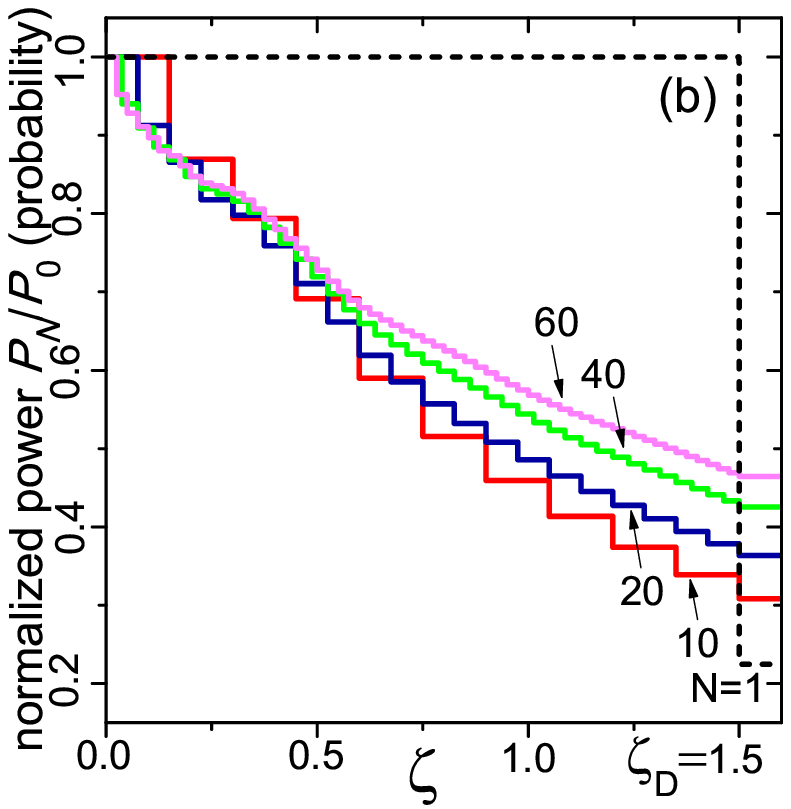}
\end{center}
\caption{\label{fig6} (Color online) For the inverse-Zeno scheme of
Fig.~\ref{fig5},  (a) normalized power $P_N/P_0$ on the detector
slit as a function of the number of intermediate, laterally
displaced slits for a fixed source-detector distance $\zeta_{\rm
D}$; (b) normalized power $P(\zeta)/P_0$ as a function of the
propagation distance $\zeta$ for different number of intermediate
slits within $\zeta_{\rm D}$. Steps are at $\zeta_n=n(\zeta_{\rm
D}/N)$, where the laterally displaced slits by $\Delta \xi_n =
n(2/N)$, $n=0,1\dots N$ for each $N$ are placed. The lowest step
of each staircase is the power that arrives at the detector in
each case.}
\end{figure}

Figure \ref{fig6}(a) depicts the light power on the detector $P_N$
as a function of $N$ for given $\zeta_{\rm D}$, showing again a
substantial rise with the number of intermediate, gradually
displaced slits, a rise that reaches the source power $P_0$ in the
limit $N\rightarrow\infty$. Figure \ref{fig6}(b) represents the
power as a function of $\zeta$ for several number of intermediate,
displaced slits within $\zeta_{\rm D}=1.5$. As described in
relation to Fig.~\ref{fig3}, light spreads between slits but
preserves its power, experiencing therefore sudden drops as light
impinges the intermediate slits. However, as in the Zeno effect,
increasing the number of steps down results in a higher last step,
that a higher detected power. In the analogous quantum mechanical
system, the particle can be said to be guided to the desired
region of space by repeatedly asking the particle if it is
gradually displaced toward that region. As in the Zeno effect,
this inverse-Zeno effect is preceded by an inverse-anti-Zeno
effect when  the source-detector distance allows low enough
frequency of the measurements.

\section{Theory}\label{THEORY}

To simplify the notation and emphasize the parallelism with
quantum mechanics, we rewrite the Fresnel integral in Eq.~(\ref{FRESNEL}) as
\begin{equation}
\psi_1(\xi)= U(\zeta)\psi_0(\xi)\,,
\end{equation}
where $\psi_0(\xi)=T(\xi)\psi(\xi)$ is the initial localized state
due to truncation with the aperture function
\begin{equation}
T(\xi)=\left\{\begin{array}{ll}1 & \mbox{if}  \quad |\xi|\le 1 \\
0 & \mbox{otherwise} \end{array}\right.\, ,
\end{equation}
and where
$U(\zeta) = \exp(-iH\zeta)$ is the evolution operator,
with $H=-\frac{1}{2}\partial^2/\partial \xi^2$ the Hamiltonian.
Setting
\begin{equation}
P_0=\int d\xi |\psi_0(\xi)|^2 =1\, ,
\end{equation}
all expressions below hold equally for the quantum probability or
the light power normalized to the source power.

\subsection{One-dimensional Zeno effect}

In a standard Zeno scheme, measurements check whether the evolved
state $U(\zeta)|\psi_0\rangle$ is in the initial localized state
$|\psi_0\rangle$. Measurement  is described by applying the
projector $|\psi_0\rangle\langle\psi_0|$ to
$U(\zeta)|\psi_0\rangle$, so that the emerging state is
$|\psi_0\rangle\langle \psi_0|U(\zeta)|\psi_0\rangle$ and the
probability of finding the initial state is given by
$|\langle\psi_0|U(\zeta)|\psi_0\rangle|^2$.  Applying this scheme
to $N$ measurements evenly spaced by $\zeta_{\rm D}/N$ in
$\zeta_{\rm D}$, the probability of finding the initial state
$|\psi_0\rangle$ in the last measurement is $P_{N}^{(S)}=|\langle
\psi_0|U(\zeta_{\rm D}/N)|\psi_0\rangle|^{2N}$, or
\begin{equation}\label{PROBS}
P_{N}^{(S)}=\left|\int d\xi \psi_0^{\star}(\xi) U(\zeta_{\rm
D}/N)\psi_0(\xi)\right|^{2N}\, ,
\end{equation}
where the superscript $(S)$ stands for  ``standard'' scheme.  The
probability is the product of individual probabilities because the
state is reset, except for its amplitude, to the initial state in
each measurement. In the standard Zeno, $U$ is usually approached
by $U =1+i H (\zeta_{\rm D}/N)+ (1/2) H^2(\zeta_{\rm D}/N)^2$ for
large-enough $N$. Further, in case that
$\langle\psi_0|H|\psi_0\rangle$ and
$\langle\psi_0|H^2|\psi_0\rangle$ are well-defined,
$P_N^{(S)}\simeq \left[1 - \left(\zeta_{\rm D}/N\right)^2(\Delta
H)^2_{\psi_0}\right]^N\rightarrow 1$ as $N\rightarrow\infty$ is
obtained, meaning that the state is $|\psi_0\rangle$ in the limit
of infinitely frequent measurements.

\subsection{Zeno effect within a subspace in wave-packet spreading}

In Sec.~\ref{NUM} we instead have checked if the position of the
quantum particle remains in the initial space domain $[-1,+1]$.
A measurement is then described by applying the projector
$\int_{-1}^{+1}d\xi |\xi\rangle \langle \xi|$, the
state after the measurement then being $\int_{-1}^{+1}d\xi |\xi\rangle \langle
\xi|U(\zeta)|\psi_0\rangle$, or $T(\xi)U(\zeta)\psi_0(\xi)$ in
position representation. Accordingly, the state after $N$
measurements of position within $\zeta_{\rm D}$ is
\begin{equation}\label{STATE}
\psi_N(\xi)=T(\xi)U(\zeta_{\rm D}/N)\cdots T(\xi)U(\zeta_{\rm D}/N)\psi_0(\xi)\, ,
\end{equation}
and the probability of finding the particle in $[-1,+1]$ is
\begin{equation}\label{P}
P_N= \int d\xi |T(\xi)U(\zeta_{\rm D}/N)\cdots T(\xi)U(\zeta_{\rm D}/N)\psi_0(\xi)|^2\, .
\end{equation}
As explained, this quantum dynamics is analogous to that of
repeated diffraction in a distance $\zeta_{\rm D}$, $P_N$ then
meaning the power in the $N$th slit normalized to the power on the
source slit. Generally, it is not possible to factorize Eq.~(\ref{P})
as in the standard Zeno scheme, since the state is not
reset to the initial one in each measurement, and other approaches
must be pursued to explain the Zeno and anti-Zeno
effects described in Sec \ref{NUM}.

\subsubsection{Fraunhofer regime}

Let us first analyze the anti-Zeno effect. Suppose that
$\zeta_{\rm D}$ is large enough, and $N$ is small enough that the
slit-to-slit distance $\zeta_{\rm D}/N\gg 1$. For the input
square wave $\psi_0(\xi)=T(\xi)/\sqrt{2}$ considered in the
preceding section, we have
\begin{equation}
 U(\zeta_D/N)\psi_0(\xi) \simeq \frac{e^{i\xi^2N/2\zeta_{\rm D}}}{\sqrt{2}}
  \left(\frac{2N}{\pi i \zeta_{\rm D}}\right)^{1/2}
  \frac{\sin(\xi N/\zeta_{\rm D})}{\xi N/\zeta_{\rm D}}\, ,
\end{equation}
where Fraunhofer diffraction has been used. Well within the
Fraunhofer region ($\zeta_{\rm D}/N\gg 1$), this pattern can be
regarded as approximately  uniform in the limited region
$[-1,+1]$, and thus we write
\begin{equation}
\psi_1(\xi)=T(\xi)U(\zeta_{\rm D}/N)\psi_0(\xi) \approx
\frac{T(\xi)}{\sqrt{2}}\left(\frac{2N}{\pi i\zeta_{\rm
D}}\right)^{1/2}
\end{equation}
for the wave just after the first slit, with power $(2N/\pi
\zeta_{\rm D})$.
Within this approximation, we can repeatedly apply the operator
$T(\xi)U(\zeta_{\rm D}/N)$ to obtain similar expressions for the wave $\psi_n(\xi)$ just
after the $n$th slit, and its power as $(2N/\pi \zeta_{\rm D})^n$. In
particular, the wave on the detector is
\begin{equation}
\psi_{N}(\xi)\approx \frac{T(\xi)}{\sqrt{2}}\left(\frac{2N}{\pi i
\zeta_{\rm D}}\right)^{N/2}\, ,
\end{equation}
and the measured power is
\begin{equation}\label{PROBANTI}
P_N\approx\left(\frac{2N}{\pi \zeta_{\rm D}}\right)^N\, .
\end{equation}
Given $\zeta_{\rm D}$, this expression is expected to be
approximately valid for $N\ll \zeta_{\rm D}$. The value of $P_N$
given by Eq.~(\ref{PROBANTI}) is seen to be a decreasing function
of $N$ for $N\ll \zeta_{\rm D}$ [Fig.~\ref{fig4}(a), dashed
curves] that provides an accurate description of the anti-Zeno
effect in wave-packet spreading. The power loss by repeated
factors $(2N/\pi \zeta_{\rm D})$ gives a good description of the
actual power loss as light impinges each intermediate slit during
its propagation from the source  to the detector
[Fig.~\ref{fig4}(b), dashed curves].

\subsubsection{Fresnel regime}

The anti-Zeno effect disappears when the slit-to-slit distance is
not in the Fraunhofer region. With increasing number of slits,
diffraction can act a sufficient number of times to shape a
diffraction wave mode, whose much lower diffraction losses can
explain the Zeno effect described in Sec.~\ref{NUM}. As originally
studied for plane, two-mirror (Fabry-Perot) resonators \cite{FL},
a diffraction mode self-reproduces upon propagation from one to
the next diffracting slit (from mirror to mirror) apart from a
complex constant or eigenvalue, that is,
$\psi_{n+1}(\xi)=\gamma\psi_{n}(\xi)$. Accordingly, the power
varies as $P_{n+1}=|\gamma|^2 P_n\equiv (1-\delta)P_n$, where
$\delta$ is the fractional power loss per slit (i.e., per mirror
reflection). The shape of a diffraction wave mode and its
fractional power loss depend on the slit spacing (mirror distance)
$L$. For the fundamental diffraction mode the fractional power
loss can be approximated by $\delta \approx 0.12 N_F^{-3/2}$ if the
resonator Fresnel number $N_F=a^2/(\lambda L)$ is greater than
unity \cite{FL}.

\begin{figure}[t]
\begin{center}
\includegraphics[width=4cm]{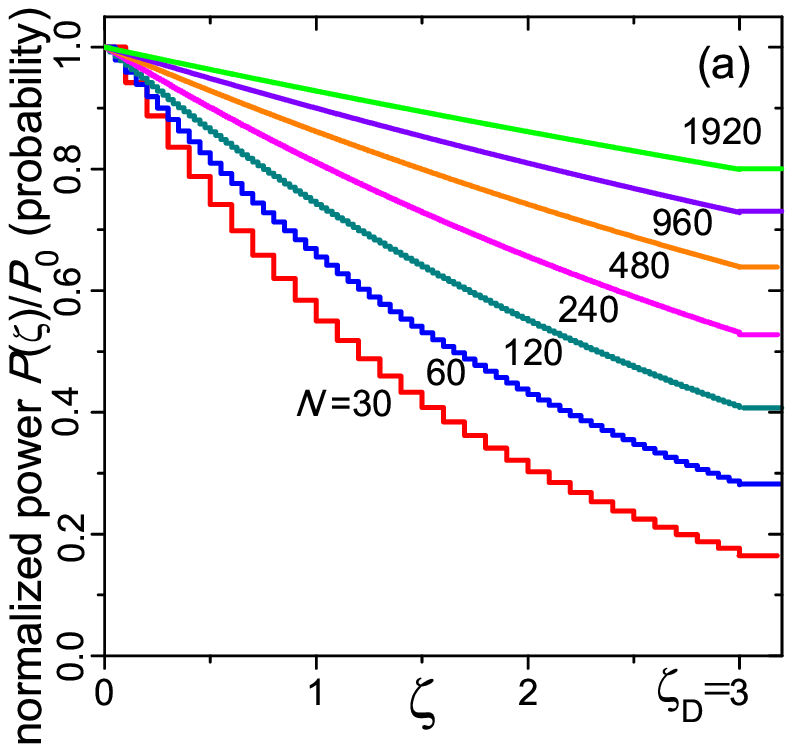}
\hspace*{0.3cm}\includegraphics[width=4cm]{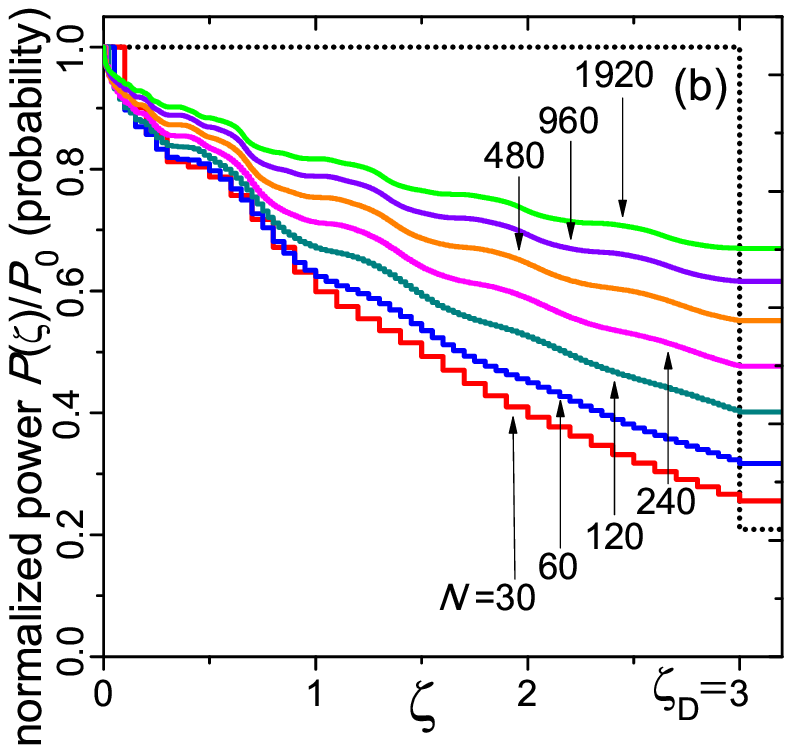}
\end{center}
\caption{\label{fig7} (Color online) Normalized power $P(\zeta)/P_0$ as a
function of the propagation distance $\zeta$ for different large
numbers of intermediate slits within $\zeta_{\rm D}=3.0$ (a) as
predicted by the model of diffraction modes, and (b) evaluated as
in Fig.~\ref{fig3}  starting with uniform illumination.}
\end{figure}

\begin{figure}[t]
\begin{center}
\includegraphics[width=5.5cm]{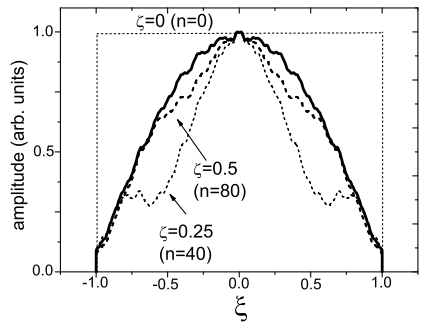}
\end{center}
\caption{\label{fig8} Solid curve: amplitude $|\psi(\xi)|$ of the
fundamental diffraction mode for $N_F\simeq N/(2\pi\zeta_{\rm
D})\simeq 25.5$, corresponding to $N=480$ slits in $\zeta_{ \rm
D}=3$. Dashed curves: amplitudes $|\psi_n(\xi)|$ on the
intermediate slits $n=0,40,80$ showing the gradual formation of
the fundamental diffraction mode explaining the Zeno effect. For better comparison, all peak
amplitudes are normalized to unity.}
\end{figure}

For a large-enough number of intermediate slits between the source
and the detector, we can assume that the diffraction mode
corresponding to the slit spacing propagates from slit to slit
and apply the above theory. In our case, the slit spacing is $L=
z_ {\rm D}/N$, the Fresnel number is $N_F=N a^2/\lambda z_{\rm
D}$, or, in our dimensionless variables, $N_F=N/2\pi\zeta_{\rm
D}$, and the fractional power loss per slit becomes
$\delta=b(\zeta_{\rm D}/N)^{3/2}$ [$b=0.12(2\pi)^{3/2}\simeq
1.8899$]. Thus, as the number of slits $N$ increases, the Fresnel
number increases proportionally, but the losses per slit decrease
as the faster rate $N^{-3/2}$, which results in a higher power on
the detector. More precisely, the variation of the power from slit
to slit is given by $P_{n+1}-P_{n}=-b(\zeta_{\rm D}/N)^{3/2}P_n$,
which leads, for small slit spacing, to the exponential decay
\begin{equation}\label{modespower}
P_n\simeq P_0\exp\left[-b \left(\frac{\zeta_{\rm
D}}{N}\right)^{1/2}\zeta_n \right]\, ,
\end{equation}
with a ``life-time'' growing as $N^{1/2}$. In particular, the power
on the detector $P_N=P_0\exp\left(-b\zeta_{\rm
D}^{3/2}/N^{1/2}\right)$  also increases as $N^{1/2}$, approaching
$P_0=1$ as $N\rightarrow\infty$. The exponentially decaying steps
of the power for different, large numbers of intermediate slits,
as predicted by Eq.~(\ref{modespower}), are represented in
Fig.~\ref{fig7}(a) in order to show that this simple model
reproduces the mechanism of the actual Zeno effect, represented in
Fig.~\ref{fig7}(b). These two figures differ quantitatively
because in (a) the fundamental diffraction mode is assumed to
propagate from the beginning, while in (b) the diffraction mode is
gradually formed from the input uniform illumination. Figure
\ref{fig8} shows the gradual formation of the diffraction mode
(thick solid curve) as the input square wave is diffracted by the
successive slits for $N=480$ slits in $\zeta_{\rm D}=3$,
corresponding to a slit spacing $0.00625$ and a Fresnel number
$N_F\simeq 25.5$ (as an example, for visible light at
$\lambda=600$~nm and a slit of width $2a=1$~mm, three diffraction
lengths would be 7854~mm and the slit spacing would be about
$16.36$~mm). The process of mode formation from the uniform
illumination causes the decays in Fig.~\ref{fig7}(b) to be
initially faster than exponential, but they are moderated to the
same exponential decays as in Fig.~\ref{fig7}(a) at distances
where the wave mode is substantially formed. The result is a
slightly less pronounced Zeno effect in the detected power
compared to the prediction of the simple model starting with
diffraction modes. Also, the uniform illumination excites spurious
higher-order diffraction modes. Since their diffraction losses are
higher, the oscillations observed in Fig.~\ref{fig7}(b) and
caused by interference with the fundamental mode, attenuate with
propagation distance.

The inverse-Zeno effect with misaligned slits can be understood
from the diffraction mode theory in a similar way as the Zeno
effect. The only significant difference is that the diffraction
wave mode that builds up by repeated diffraction and the
corresponding losses per slit, are those of a misaligned,
two-mirror laser resonator, as described in specific studies on
the effects of misalignment on these resonators \cite{FL}.

\section{Conclusions}
\label{conc}

We have analyzed a simple and feasible scheme displaying Zeno, anti-Zeno,
and inverse-Zeno effects, valid in quantum mechanics and classical wave optics.
The classical wave optics scheme is particularly simple since observation of the system
is achieved just by inserting slits between a light source and a light-power
detector.

We have shown that the occurrence of Zeno or anti-Zeno effect
depends on the separation between the source and the detector and
the number of intermediate slits. The anti-Zeno effect seems more
intuitive: Adding diffracting slits increases diffraction and
holds for large enough separation between slits (i.e., less
frequent measurements). This separation is close to the
diffraction length (which is determined by slit size and light
wavelength). On the other hand, the Zeno effect holds for smaller
separation between slits (i.e., more frequent measurements) and
corresponds to the rather counterintuitive behavior that
diffraction is inhibited by placing more and more diffracting
slits between the source slit and the detector. The slit
separation for the transition between Zeno and anti-Zeno effects
is the analog of the so-called Zeno time
\cite{Lane83,Schieve89,Kofman96,Alfredo98,Maniscalco06,Francica10}.

We have related the Zeno effect to the diffraction mode theory in
Fabry-Perot resonators. The total power losses during passage
through tightly space slits are less than the total losses through
widely spaced slits within the same distance. This is because a
diffraction mode of the equivalent resonator tends to be formed by
the many diffractions and because the mode losses are lower as the
slit spacing diminishes. In this regard, the limit of a continuous
of measurements in the quantum domain has been considered in Ref.~\cite{Fakki,rev},
where it is shown that the wave packet evolves as in
an infinitely deep well potential. In classical wave optics, such an
ideal limit might be regarded as a perfect conductor waveguide
forcing the vanishing of the electric field at its walls, avoiding
diffraction losses by expelling the field away from the walls.

\section*{Acknowledgments}

Support from the Ministerio de Ciencia e Innovaci\'{o}n (Spain)
under Projects FIS2010-22082 (M. A. P., I. G., A. S. S.),
FIS2008-01267 (A. L.), and FIS2010-18132 (A. S. S.) is
acknowledged. A. L. also acknowledges support from Project
QUITEMAD S2009-ESP-1594, from the Consejer\'{\i}a de Educaci\'{o}n
de la Comunidad de Madrid. A. S. S. would also like to thank the
Ministerio de Ciencia e Innovaci\'{o}n (Spain) for a ``Ram\'on y
Cajal'' Grant.

\end{document}